\documentstyle[prl,aps]{revtex}
\begin{document}
\input{epsf}
\draft

\title{%
\leftline{\small KUCP-0095}
\leftline{\small KUNS-1390 HE(TH)96/03}
\leftline{\small OHU-PH 9606}
\leftline{\small HEP-TH/9606159}
Fake Instability in the Euclidean Formalism}
\author{Hideaki Aoyama\cite{ackaoy}}
\address{Faculty of Integrated Human Studies,
Kyoto University, Kyoto 606-01, Japan\\
{\sl aoyama@phys.h.kyoto-u.ac.jp}}
\author{Toshiyuki Harano\cite{ackgakusin}}
\address{Faculty of Science, Kyoto University, Kyoto 606-01, Japan\\
{\sl harano@gauge.scphys.kyoto-u.ac.jp}}
\author{Hisashi Kikuchi\cite{ackkiku}}
\address{Ohu University, Koriyama 963, Japan\\
{\sl kikuchi@yukawa.kyoto-u.ac.jp}}
\author{Masatoshi Sato\cite{ackgakusin}}
\address{Faculty of Science, Kyoto University, Kyoto 606-01, Japan\\
{\sl msato@gauge.scphys.kyoto-u.ac.jp}}
\author{Shinya Wada\cite{ackgakusin}}
\address{Graduate School of Human and Environmental Studies,
Kyoto University, Kyoto 606-01, Japan\\
{\sl shinya@phys.h.kyoto-u.ac.jp}}
\maketitle

\vspace{4mm}
\centerline{{\bf Abstract}}
\vspace{2mm}
\begin{abstract}
We study the path-integral formalism in the imaginary-time
to show its validity in a case with a metastable ground state.
The well-known method based on the bounce solution
leads to the imaginary part of the energy even for
a state that is only metastable and has a simple oscillating behavior
instead of decaying.
Although this has been argued to be the failure of the
Euclidean formalism, we show that proper account of the
global structure of the path-space leads to a valid
expression for the energy spectrum, without the imaginary part.
For this purpose we use the proper valley method to
find a new type of instanton-like configuration, the
``valley instantons''.
Although valley instantons are not the solutions of equation of motion,
they have dominant contribution to the functional integration.
A dilute-gas approximation for the valley instantons
is shown to lead to the energy formula.
This method extends the well-known imaginary-time formalism
so that it can take into account the global behavior of the
theory.
\end{abstract}


\section{Introduction}
The imaginary-time formalism has been successful for studies of the
various
quantum tunneling phenomena in the semi-classical regime.
This is because of the existence of the solution of the Euclidean equation
of motion, around which we could evaluate the relevant
functional integration in Gaussian and higher order approximations.
In the systems with finite degrees of freedom, this is known to
lead to WKB results.

More specifically, in case when there are perturbatively degenerate
vacua (or ground states), the instanton calculation
takes into account the tunneling between them and gives the
nonperturbative contribution to the energy splitting \cite{polya,clec}.
On the other hand, when a perturbative vacuum is unstable due to the
tunneling to lower energy states, the bounce
solution \cite{clec,coleman,cc} leads to
the imaginary part of the energy, thus the decay rate.
This formalism provides us with a good calculational
tool, valid in a wide range of physical systems.

\begin{figure}
\centerline{\epsfxsize=10cm\epsfbox{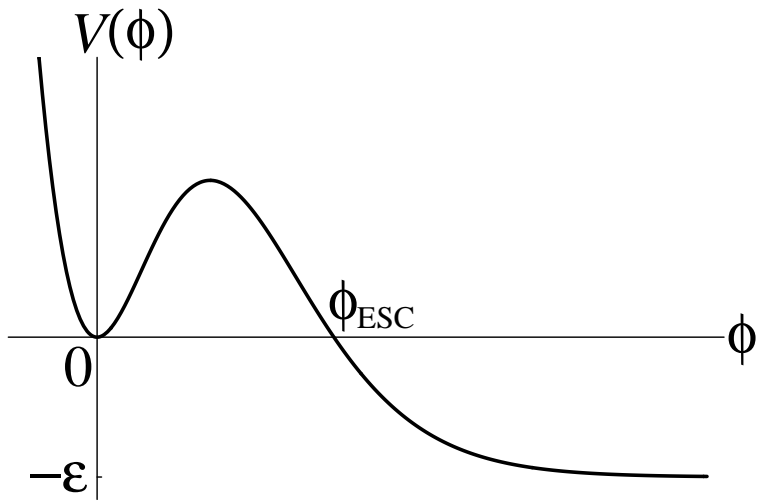}}
\caption{A potential that is flat in the asymptotic direction,
$V(\infty) = -\epsilon$.}
\label{fig:potflat}
\end{figure}

One subtle  feature of these calculations can be seen
for a one-dimensional quantum mechanical model
with a potential, $V(\phi)$, illustrated in Fig.\ref{fig:potflat}.
(We denote the coordinate by $\phi$ in this letter.)
If one restricts the wave functions to have only the outgoing component
at $\phi \gg \phi_{\rm ESC}$,  the hermiticity of the Hamiltonian
is violated and the energy eigenvalues become complex; this imaginary part
is a direct consequence of the instability of the localized wave packet at
$\phi \sim 0$.
In the imaginary-time formalism, this complex energy is thought to be
evaluated by an analytic continuation of the  ill-defined divergent Gaussian
integral over
the negative mode direction at the bounce solution \cite{clec,coleman,cc}.

\begin{figure}
\centerline{\epsfysize=9cm\epsfbox{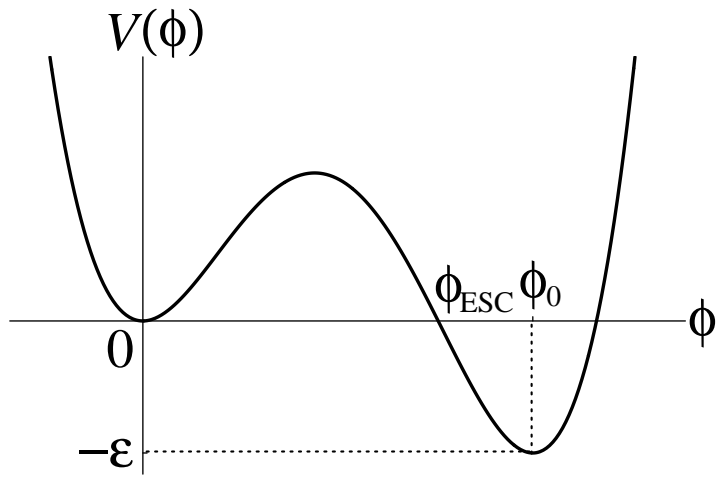}}
\caption{An asymmetric double-well potential, in which $\phi=0$ is
only metastable.}
\label{fig:potdw}
\end{figure}

We may then ask ourselves what are the eigenvalues for an unbalanced
double-well potential illustrated in Fig.\ref{fig:potdw} \cite{boya}.
(In order to distinguish this case from the previous one,
we call this `metastable' case and the previous one `unstable'.)
As far as we stick to the classical solution and take into account
only the contribution
of its infinitesimal neighborhood for the path integral, the result is the
same as the unstable case;  we have a similar bounce solution with a
negative mode and obtain complex eigenvalues.

This imaginary part is obviously a fake.
The perturbative ground state in the left well is only metastable in the
sense that any wave packet that tunnels to the right well oscillates between
the two wells.
This oscillation generates the splitting behavior
similar to the degenerate case, $\epsilon =0$, but not to the decay rate.
In other words, we can only take wave functions with decaying exponential
at $\phi \gg \phi_0$ for the potential in Fig.\ref{fig:potdw}
and the hermiticity  of the Hamiltonian cannot be violated.

It was claimed that Euclidean formalism is doomed due to the fact
that it only uses the information around the infinitesimal neighborhood
of the solutions\cite{boya}.
As an alternative, the complex-time method was proposed,
in which the real-time part of the classical trajectory
takes into account the correct boundary condition
for $\phi \ge \phi = \phi_{\rm ESC}$.
Although this is a very exciting development by itself\cite{cn,ah},
it is not clear whether it should replace the Euclidean formalism.

In this letter we show that the proper treatment of the imaginary-time
path integral leads us to the correct behavior of the energy
eigenvalues for the metastable potential.
The essence of the improvement is the use of the
{\em proper valley method},  which was developed independently by
Silvestrov \cite{silv} and two of the present authors \cite{ak}.
It teaches us how to enlarge the set of background configurations
besides the classical solutions in order to take into account
the global structure of the functional space.
Using this method,  we construct ``valley instanton'',
which should replace the bounce solution.
Interestingly, it has a zero mode and this expedites the calculation
of its determinant and Jacobean as was the case of the instanton.
It is a well-localized configuration with respect to the imaginary
time and their dilute gas sum generates the reasonable
energy corrections of the lowest states instead of the imaginary part.
We will also show that it converges analytically to the instanton
in the limit of \(\epsilon \rightarrow 0\) and all the results
reproduce the well-known instanton results.

\section{The valley bounce and the valley instanton}

Consider the quantum mechanical system with the following Euclidean
action;
$$S_{\rm E}[\phi] = \int d\tau  \left( {1 \over 2}
\left({d\phi \over d\tau}\right)^2 + V(\phi) \right)\; ,
\label{eq:action}$$
with the potential,
\begin{equation}
V (\phi) = {1 \over 2} \phi^2 \left(1 - g\phi\right)^2
- \epsilon (4 g^3 \phi^3 - 3 g^4 \phi^4) ,
\label{eq:potone}
\end{equation}
where the coupling constants $g$ and $\epsilon$ are positive.
The potential (\ref{eq:potone}) has a local minimum, $V(0) = 0$ and
a global minimum at $\phi_0 = 1/g$, where $V(1/g) = -\epsilon$.
(Fig.\ref{fig:potdw} is plotted for $g=0.3$ and $\epsilon=0.25$.)
The potential $V(\phi)$ in Eq.(\ref{eq:potone}) is a
canonical form of quartic potentials with two minima, since
any such potential can be cast into this form by suitable linear
transformations on $\phi$ and $\tau$. In this sense the following
analysis is quite a general one.
In the following, we consider the cases with small coupling $g \ll 1$,
but not necessarily small $\epsilon$.

As mentioned in the introduction, the naive application of the
semi-classical approximation around the bounce solution of this potential
leads us to the fake imaginary part of the energy eigenvalues.
In order to circumvent this, we construct the different type of
configurations around which we expand the action.
Such configurations are most straightforwardly identified by the valley
methods\cite{yung}, or more specifically,  by
the proper valley method\cite{silv,ak}.
The latter has many advantages over the previous one.
It has been applied to the Borel summability problem\cite{kiku},
induced bubble nucleation problem\cite{aw}, the
instanton in gauge-Higgs system\cite{ahsw},
and the baryon number violation problem\cite{hs} successfully.

For the current model,  the advantage of using the proper valley
method is that we can see  how the
action behaves if we go along  the  negative mode direction at the bounce,
the direction most important to evaluate  the  path integral correctly.
As we will see below,  it reaches to $\phi \sim \phi_0$, and see
the behavior of the potential around there.

The proper valley configurations are given by the following new valley
equations;
\begin{eqnarray}
-\partial_\tau^2 \phi + V'(\phi) &= F, \label{eq:nvone}\\
\left( -\partial_\tau^2  + V''(\phi)\right) F &= \lambda F.
\label{eq:nvtwo}
\end{eqnarray}
For $F=0$, this set of equations reduces to the ordinary equation of motion.
Thus, any solution of the equation of motion is the solution of the
new valley equations.
Otherwise, eliminating $F$ one finds that the left-hand side
of Eq.(\ref{eq:nvone}), which is $\delta S / \delta \phi$,
is the eigenvector of the $\delta^2 S / \delta \phi \delta \phi (\equiv D)$
with the eigenvalue $\lambda$.
The general solutions can be parametrized by the eigenvalue $\lambda$,
or any arbitrary function $\alpha$ of $\lambda$.
We  denote this one-parameter family of the solutions, the valley trajectory,
by $\phi_\alpha (\tau)$.
The contribution of such a configuration to the vacuum transition amplitude,
\begin{equation}
\langle \phi=0, \tau=+\infty |\phi=0, \tau=-\infty \rangle
= {\cal N} \int_{\phi(\pm\infty)=0} {\cal D}\phi e^{-S[\phi]},
\label{eq:vvamp}
\end{equation}
is evaluated around this valley by inserting the following triviality;
\begin{equation}
1 = \Delta\int d\alpha
\delta\left(
        \int d\tau (\phi - \phi_\alpha){\delta S \over \delta\phi_\alpha}
\right)\; .
\end{equation}
This forces the expansion of $\phi-\phi_\alpha$
in the subspace orthogonal to $\delta S / \delta \phi$,
enabling the Gaussian integration without the linear terms.
At the leading order of $\hbar$, we obtain;
\begin{equation}
\langle \phi=0, \tau=+\infty |\phi=0, \tau=-\infty \rangle
= {\cal N} \int d\alpha {1 \over \sqrt{\det D^\prime}} J_\alpha
e^{-S[\phi_\alpha]},
\label{eq:vvampalpha}
\end{equation}
where $\det D^\prime$ is the usual determinant less the
eigenvalue $\lambda$ in the new valley equation (\ref{eq:nvtwo}) and
\begin{equation}
J_\alpha \equiv {\displaystyle \int d\tau {d \phi_\alpha \over d\alpha}F
\over \displaystyle\sqrt{\int d\tau F^2}}
= {\displaystyle{d S[\phi_\alpha] \over d\alpha} \over
\displaystyle\sqrt{\int d\tau F^2}}.
\label{eq:jacob}
\end{equation}
This way we are able to perform the integral for  the negative mode
direction on a more rigid basis than the subtle Gaussian integral.
The factor $J_\alpha$ is the Jacobian for this change of the integration
variable.
We also note that Eq.(\ref{eq:vvampalpha}) is apparently invariant under
any local reparametrization in $\alpha$.

\begin{figure}
\centerline{\epsfysize=10cm\epsfbox{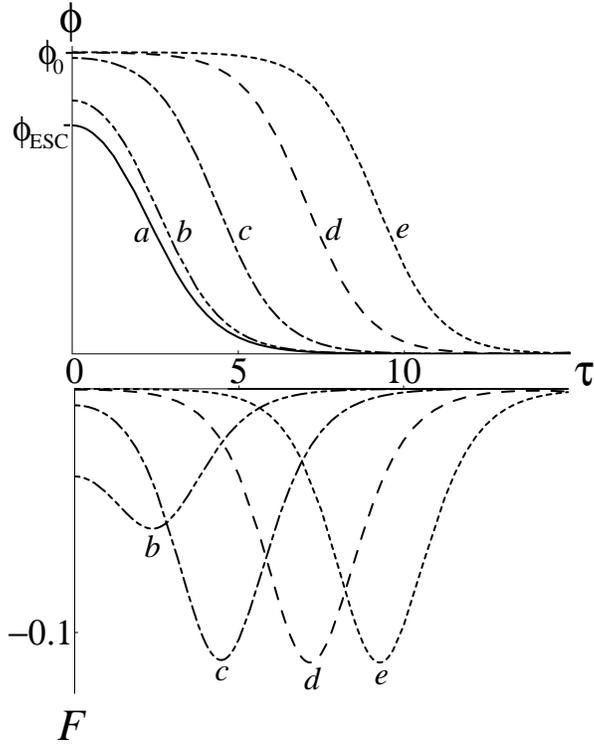}}
\vspace{3mm}
\caption{Valley bounce solutions $(\phi(\tau), F(\tau))$ of the new
valley equations.
Center of all of the configurations are chosen to be at the origin, $\tau=0$,
around which the solutions are symmetric.
The solid line $a$ in the upper figure is the usual bounce solution, which has
$F(\tau)=0$.
The other lines, $b$--$e$, are unique to the new valley equations.}
\label{fig:conf}
\end{figure}

\begin{figure}
\centerline{\epsfysize=9cm\epsfbox{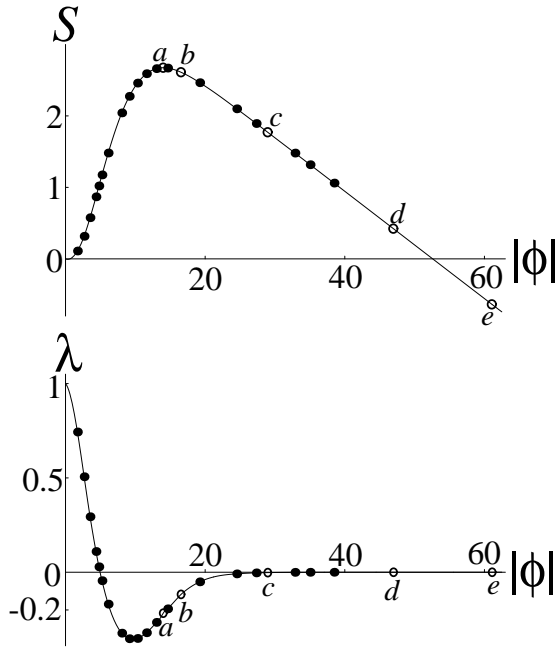}}
\vspace{3mm}
\caption{The values of the action, $S$, and the eigenvalue, $\lambda$, of the
valley bounces.
The peak of the action is given by the bounce solution, the line $a$ in
Fig.3. The points corresponding to the valley bounces in Fig.3 are
plotted with circle.
The solid lines are drawn as the guide for eyes.
The valley parameter is chosen to be $|\phi| \equiv  \int d\tau \phi$.}
\label{fig:action}
\end{figure}

We have carried out the numerical analysis and obtained the solutions
plotted in Fig.\ref{fig:conf}.
This is a simple extension of the previous analysis for the meta-stability
problem in a quantum field theory  by two of the current authors, H.~A.~and
S.~W.~\cite{aw}.
The solid line, $a$, is the bounce solution of the equation of motion.
The rest have $F \ne 0$ do not satisfy the equation of motion.
We call these solutions (including the bounce solution) ``valley bounce".
The values of the action, $S$, and the eigenvalue, $\lambda$,
for the valley bounces are plotted in Fig.\ref{fig:action}.
The bounce solution lies at the top of the line of the action
in Fig.\ref{fig:action}, corresponding to the fact that it has
a negative eigenvalue.
This negative eigenvalue can be read from the corresponding
point of the plot of $\lambda$ in the lower half of Fig.\ref{fig:action}.
The most notable feature is that the large size valley bounces
have a clean interior, where $\phi = \phi_0$ and $F = 0$.
(This property is shared by the higher dimensional configurations,
{\it i.e.}, the valley bubbles.)
The effect of this is apparent in the behavior of the action
in Fig.\ref{fig:conf};
the action decreases linearly with large $|\phi|$,
which is proportional to the size of the valley bounce.
Thus, the valley bounces provide a natural way for the evaluation of
the contribution of large regions of the true minimum $\phi_0$.

In the large valley bounces, $c$, $d$, and $e$ in Fig.\ref{fig:conf},
we notice that the shape of the wall, {\it i.e.,}
the transition region from $\phi \sim \phi_0$ to $\phi \sim 0$,
is almost identical to each other.
That is, they overlap with each other very well when translated in $\tau$.
Thus these large valley bounces can be approximated
by simply connecting the walls by a flat region, $\phi=\phi_0$, at
various separation.
The shape of the wall can be most readily identified
when the size ($|\phi| \equiv \int d\tau \phi $) of
the valley bounce becomes $\infty$.
In this limit, the wall is simply a localized transition from
$\phi = 0$ to $\phi = \phi_0$ (or vise-versa).
Such a solution is an analogue of the instanton (or anti-instanton).
The difference is that now it is not a solution of equation of motion,
but is a solution of the new valley equation.
This kind of configuration is called the ``valley instanton" \cite{ahsw}.
In the following, we evaluate the properties of the valley instantons,
in order to approximate the large valley bounces by a pair of
a valley instanton ($\phi=0 \rightarrow \phi_0$) and a valley anti-instanton
($\phi=\phi_0 \rightarrow 0$).

Since we define the valley instantons at the large-size limit,
$|\phi| \rightarrow \infty$, of the valley bounce,
the plot of the eigenvalue in Fig.\ref{fig:conf} implies that
the eigenvalue $\lambda$ of the valley instanton is exactly zero.
This is not a trivial property.  A solution of equation of motion
is guaranteed to have zero modes corresponding to its symmetry
transformation, such as a time translation.
Arbitrary background configurations do not have this property in general.
However, we can prove the existence of the zero mode as follows:
Take a derivative of Eq.(\ref{eq:nvone}) with respect to $\tau$,
multiply $F$, and integrate over $\tau$.
After partial integrations (which surface terms vanish), we then find that
\begin{equation}
\lambda \int_{-\infty}^\infty F \dot{\phi} d\tau
= \int_{-\infty}^\infty F \dot{F} d\tau \; .
\end{equation}
The integral in the left-hand side is generally non-zero
due to the boundary conditions of the valley instantons.
This can be seen from the behavior of $\phi$ and $F$ in the walls
in Fig.\ref{fig:conf}.
Since the right-hand side is zero, we find that $\lambda = 0$.
(For the valley bounces, the integral in the left-hand side is zero
since $\phi(\tau)$ and $F(\tau)$ are even functions.
Therefore, $\lambda \ne 0$ is allowed.)

We have carried out numerical investigation of the
valley instanton with $\lambda=0$ and have successfully obtained
the solutions, which have turned out to be almost identical
to the wall regions in Fig.\ref{fig:conf}.
Although we know of no exact analytical expression of the valley instanton,
it can be constructed analytically for small $g^2$
in a manner used in the construction of the constrained
instanton \cite{affleck}
as well as other types of the valley instantons \cite{ahsw}.
Consider the valley instanton in $\tau \in [-T/2, T/2]$ ($T \gg 1$).
We define its central coordinate to be at the origin,
$\tau = 0$, by $\phi(0)=1/2g$.
The naive perturbation in $\epsilon g^2$ yields the following perturbative
valley instanton solution;
\begin{eqnarray}
\phi &=& \phi_0^I + 3\epsilon g^2 \tau \dot{\phi}_0^{I}
+ O((\epsilon g^2)^2) ,\nonumber\\
&&{} \label{eq:pertsol} \\
F &=& - 6 \epsilon g^2 \dot{\phi}_0^I
+ 36 \epsilon^2 g^5 \tau \phi_0^I \dot{\phi}_0^I + O((\epsilon g^2)^3),
\nonumber
\end{eqnarray}
where $\phi_0^I$ is the instanton solution for $\epsilon=0$;
\begin{equation}
\phi_0^I = {1 \over g} {1 \over 1 + e^{- \tau}}.
\end{equation}
{}From Eq.(\ref{eq:pertsol}), it is apparent that this naive perturbation
is valid only in the region close to the instanton center,
$|\tau| \ll 1/(\epsilon g^2 )$.
On the other hand, in the asymptotic regions, $\tau \rightarrow \pm \infty$,
we linearize the new valley equation and find the general solutions
valid for $|\tau| \gg 1$.
Coefficients of the general solutions are fixed by matching it with
the inner solution Eq.(\ref{eq:pertsol}) in the intermediate region,
$1 \ll |\tau| \ll 1/ (\epsilon g^2 )$.  We have found that
this procedure can be done consistently.
The resulting asymptotic behaviors are
for $\tau \rightarrow + \infty$;
\begin{eqnarray}
\phi &\simeq& {1 \over g} \left( 1 -
\left( 1 + {3 \epsilon g^2 \over \omega_-} \tau \right)
e^{- \omega_- \tau} \right) ,\nonumber\\
&& \label{eq:asymplus}\\
F &\simeq& -6 \epsilon g e^{-\omega_- \tau}, \nonumber
\end{eqnarray}
where $\omega_-^2 \equiv V^{\prime\prime}(\phi_0) = 1 + 12 \epsilon g^2 $,
and for $\tau \rightarrow -\infty$;
\begin{eqnarray}
\phi &\simeq& {1 \over g} \left( 1 + 3 \epsilon g^2 \tau \right)
e^{\tau} ,\nonumber\\
&& \label{eq:asymminus}\\
F &\simeq& -6 \epsilon g e^\tau. \nonumber
\end{eqnarray}
This way, the valley instanton is constructed in all regions of $\tau$
for small $\epsilon$.

The action of the valley instanton is given by
$S^I=1/6g^2 + \epsilon (-T/2  + 1/2) + O(\epsilon^2)$.
This action is divided to the volume part and the
remaining (proper) part as $S^I = -\epsilon T/2 + {\tilde S}^I$.
{}From the construction above, we find that
${\tilde S}^I = 1/6g^2 + \epsilon /2 + \ldots$.
However, there is a subtlety on this point:
In the following we integrate over the position coordinate of the
instantons and anti-instantons in the dilute gas approximation.
These coordinates are originally the valley parameters ($\alpha$s) of the
valley bounces (and their central coordinates).
The $O(\epsilon)$ term in ${\tilde S}^I$ depends on the definition of
these valley parameters, since the definition of the volume ($T$)
is affected by it. Therefore, careful study of the small valley bounces
is needed to fix this term.  This term, however, has only the
nonleading contribution. Therefore, we will not pursue this
problem any further here.

The Jacobian for the instanton position is given by,
\begin{equation}
J^I={\epsilon \over \displaystyle\sqrt{\int d\tau F^2}} \; .
\label{eq:acja}
\end{equation}
Since the contribution to the integration is dominated by the
central region, the leading term of Eq.(\ref{eq:acja}) for small
$\epsilon$ is evaluated by the use of Eq.(\ref{eq:pertsol}).
The result is that $J^I = 1/\sqrt{6g^2}(1 + O(\epsilon g^2))$.
The first term is the instanton action for $\epsilon =0$.
Therefore, in the limit $\epsilon \rightarrow 0$, the Jacobian of the
valley instanton reduces to that of the ordinary instanton.

The determinant, $\det D^\prime$, can be calculated by extending
the Coleman's method \cite{clec}, in spite of the fact that the valley
instanton
is not the solution of the equation of motion.
This is due to the fact that the valley instanton
possesses the exact zero mode $F(\tau)$.
We define the asymptotic coefficients $F_\pm$ by
$F(\tau) \simeq F_{\pm} e^{\mp\omega_{\pm}\tau}$,
where for the sake of notation we introduced
$\omega_+ = V^{\prime\prime}(0)=1$.
After some calculation, we find that the ratio of the determinants
is given by the following;
\begin{equation}
{\det^\prime (-\partial_\tau^2 + V^{\prime\prime}(\phi^I))
\over \det (-\partial_\tau^2 + \omega_+^2)}
= \kappa e^{ (\omega_+ - \omega_-) (T/2 - \tau_0)},
\quad
\kappa \equiv {1 \over 2 \omega_+ \omega_- F_+ F_-} \int_{-\infty}^\infty
F^2.
\end{equation}
The exponential factor is the perturbative contribution to the
zero energy at the true minima, $\phi_0$.
The factor $\kappa$ is the `proper' instanton contribution.
{}From Eq.(\ref{eq:asymplus}) and Eq.(\ref{eq:asymminus}), we find that
$\kappa$ reduces to the ordinary instanton determinant for
$\epsilon \rightarrow 0$.

\section{Path Integral in the ``dilute-gas" valley instanton approximation}

Combining all factors evaluated in the previous section,
we find the expression of the finite time ($T$) vacuum transition
amplitude to be the following;
\begin{equation}
Z(T) \equiv
{\langle \phi=0, \tau=T | \phi=0, \tau=0\rangle
\over
\langle \phi=0, \tau=T | \phi=0, \tau=0\rangle_0}
= \sum_{n=0}^\infty \alpha^{2n} I_n ,
\label{eq:amplitude}
\end{equation}
where the amplitude $\langle \ldots \rangle_0$ is for the
harmonic oscillator with $\omega_+$,
$n$ is the number of the valley instanton pairs,
and the factor $\alpha$ is the product of the proper contributions of
the action, the determinant ratio and the Jacobian;
$\alpha=(J^I / \sqrt{\kappa}) e^{-{\tilde S}^I}$.
The actual integrations over the positions of the valley instantons are
in the factors $I_n$;
\begin{eqnarray}
I_n (T) \equiv  \cases{
1, & for $n=0$, \cr
&\cr
\displaystyle \int_0^T d\tau_{2n} \int_0^{\tau_{2n}} d\tau_{2n-1}
...
\int_0^{\tau_{2}} d\tau_{1} \,
e^{\tilde\epsilon (\tau_{2n}-\tau_{2n-1}+ ... +\tau_2 - \tau_1)},
& for $n \ge 1$, \cr}
\label{eq:idef}
\end{eqnarray}
where zero-energy contributions of the determinants are
absorbed in $\epsilon$ by
$\tilde \epsilon \equiv \epsilon - (\omega_- - \omega_+)/2$.

The infinite series can be summed by the use of the generating function
method
\cite{aq}:
{}From Eq.(\ref{eq:idef}), we find that the following differential equation
is satisfied by $Z(T)$;
\begin{equation}
Z(T)'' - \tilde \epsilon Z(T)' - \alpha^2 Z(T) =0.
\end{equation}
Also, $Z(0) = 1$, and $Z'(0) =0$.
Therefore, we find that
\begin{equation}
Z(T) = {k_+ e^{-k_- T} - k_- e^{-k_+ T} \over k_+ - k_-},
\label{eq:ztresult}
\end{equation}
where,
\begin{equation}
k_\pm \equiv - {\tilde\epsilon \over 2} \pm
\sqrt{{\tilde\epsilon^2 \over 4} + \alpha^2}.
\end{equation}
Thus we find that the energies of the two lowest states,
$E_\pm$ is given by
\begin{equation}
E_\pm  = {\omega_+ \over 2} + k_\pm =
{\omega_+ \over 2} - {\tilde\epsilon \over 2} \pm
\sqrt{{\tilde\epsilon^2 \over 4} + \alpha^2}.
\label{eq:enefi}
\end{equation}
Furthermore, from the coefficients of the respective exponents of
Eq.(\ref{eq:ztresult}), , we find that
$|\langle \phi=0 | E_\pm\rangle|^2 = \pm k_\pm / (k_+ - k_-)$
This is the main conclusion of this letter.  There appear no
fake imaginary parts in the energy spectrum.
We observe that $E_\pm$ are equal to the eigenvalues of the
following matrix;
\begin{equation}
H = \pmatrix{ \displaystyle{\omega_+ \over 2} & \alpha \cr
              \alpha & \displaystyle -\epsilon + {\omega_- \over 2}\cr} .
\label{eq:hmatrix}
\end{equation}
Furthermore, the weights of the state localized in the left well agree with
Eq.(\ref{eq:ztresult}): Denoting the eigenvectors of Eq.(\ref{eq:hmatrix})
$V_\pm$ with eigenvalues $E_\pm$,
\begin{equation}
\pmatrix{1 \cr 0} = \sqrt{k_+ \over k_+ - k_-} V_+
+ \sqrt{-k_- \over k_+ - k_-} V_-.
\end{equation}
This allows a simple explanation of the result.
The energy spectrum we obtain is the same as the two-level
system made of the perturbative ground state at $\phi=0$ and
$\phi = \phi_0$, with the tunneling matrix element $\alpha$.

Before discussing the implication of the result Eq.(\ref{eq:enefi}),
we examine the validity of the dilute gas approximation we used.
The mean size of the bounce, which is the mean distance, $R$, between
the instanton and the anti-instanton located at the right of the instanton,
can be obtained as the expectation value,  $\langle\tau_2 - \tau_1 \rangle$
in the amplitude, Eq.(\ref{eq:amplitude}) and Eq.(\ref{eq:idef}).
(Any other $\langle \tau_{2m}-\tau_{2m-1}\rangle$ with integer $m$ would
yield the same result for $T \rightarrow \infty$.)
Using the same generating function method as above, we
obtain the following;
\begin{equation}
R = {1 \over \alpha^2} \left(
{\tilde\epsilon \over 2} + \sqrt{{\tilde\epsilon^2 \over 4} + \alpha^2}
\right) .
\end{equation}
Similarly, the mean distance, $d$, between
the anti-instanton and the instanton located at the right of the
anti-instanton ($\langle \tau_3 - \tau_2 \rangle$) is given by,
\begin{equation}
d = {1 \over \alpha^2}
\left(-{\tilde\epsilon \over 2} + \sqrt{{\tilde\epsilon^2 \over 4} + \alpha^2}
\right).
\end{equation}
For $\epsilon \gg \alpha$, $d \sim 1/\tilde\epsilon$.
On the other hand, the thickness of the instanton is
O($1/\sqrt{\epsilon})$ for $\epsilon \gg 1$, which can be seen by a simple
scaling argument on the new valley equations,
Eq.(\ref{eq:nvone}) and Eq.(\ref{eq:nvtwo}).
This means that the dilute gas approximation is valid for $\tilde\epsilon <
1$.

For $\epsilon \ll \alpha$,
the energy spectrum Eq.(\ref{eq:enefi}) gives the following;
\begin{equation}
E_\pm  = {\omega_+ + \omega_- \over 4} \pm \alpha - {\epsilon \over 2}
+ O\left({\epsilon^2 \over \alpha}\right).
\end{equation}
In the limit $\epsilon \rightarrow 0$, this result reduces to
the well-known instanton result for the degenerate case.
In addition, we have the average of the perturbative zero-point energies
of the left-well ($\omega_+ / 2$) and the right-well
($\omega_-/2 -\epsilon$).
Since the wavefunction is distributed evenly at the zeroth order of the
$\epsilon$ expansion, this is the correct formula for the two lowest energy
eigenvalues.  Since the perturbative contribution to the energy splitting,
$\delta E = E_+ - E_-$, is of order $\epsilon g^4$, even a purist\cite{clec}
would retain our result, $\delta E = 2 \alpha$.
For $\alpha \ll \epsilon < 1$, Eq.(\ref{eq:enefi}) leads to,
\begin{equation}
E_+  = {\omega_+ \over 2} + {\alpha^2 \over \tilde\epsilon}
+ O\left({\alpha^4 \over \epsilon^3}\right), \quad\quad
E_-  = -\epsilon + {\omega_- \over 2} - {\alpha^2 \over \tilde\epsilon}
+ O\left({\alpha^4 \over \epsilon^3}\right).
\end{equation}
The lower energy, $E_-$ corresponds to the state almost localized in
the right well.

\section{Conclusion and Discussion}
In this letter we have applied the proper valley method to
take into account the large nonperturbative configurations
in the path integral.
We have found and constructed a new type of instanton, the valley instanton,
both numerically and analytically.
We further evaluated its action, Jacobian, and determinant
and found that these have a smooth limit to the ordinary instanton for
the degenerate case.
The dilute valley-instanton approximation
to the path integral has lead us to the energy formula Eq.(\ref{eq:enefi}).
Thus we have successfully shown that proper treatment of the imaginary
time formalism does {\it not} lead to any contradiction and in fact yields
the valid energy formula.

Although the dilute valley instanton gas approximation fails
for $\epsilon > 1$,
this does not limit the applicability of the proper valley method.
For such a case, the valley bounces on the background $\phi_0$
are expected to become important.
Such valley bounces are known to exist from the analysis
of \cite{aw} and allows us to take into account the contribution of the
configurations with $\phi < \phi_0$.

In view of the current development,
the ordinary calculation of the imaginary part of the
energy in the unstable cases (Fig.\ref{fig:potflat})
needs to be examined under the new light.
This is under way and will be reported elsewhere.

\section{Acknowledgments}
We thank our colleagues at Kyoto University for discussion
and encouragements.
One of the authors, H.~A., acknowledges the hospitality by
Bum-Hoon Lee and his colleagues at the Hangyang University (Korea)
during the Fourth Haengdang Summer workshop, where part of this work was
done.
He also thanks the organizers and the participants the JRDC Workshop at
Shonan, Japan, especially, K.~Fujikawa at University of Tokyo and M.~Ueda
at Hiroshima University for encouragements and discussions.

\newpage

\end{document}